\documentclass{emulateapj}

\slugcomment{Accepted by ApJ on 2014 January 15}
\shorttitle{Two AGNs in Merger Remnant COSMOS\,J100043.15+020637.2}
\shortauthors{Wrobel, Comerford \& Middelberg}
\begin{document}

\title{Constraints on Two Active Galactic Nuclei in the Merger
  Remnant\\ COSMOS\,J100043.15+020637.2}

\author{J. M. Wrobel\altaffilmark{1,2},
  J. M. Comerford\altaffilmark{3}, and E. Middelberg\altaffilmark{4}}

\altaffiltext{1}{National Radio Astronomy Observatory, P.O. Box O,
  Socorro, NM 87801; jwrobel@nrao.edu}

\altaffiltext{2}{The National Radio Astronomy Observatory (NRAO) is a
  facility of the National Science Foundation, operated under
  cooperative agreement by Associated Universities, Inc.}

\altaffiltext{3}{Department of Astrophysical and Planetary Sciences,
  Universityof Colorado, Boulder, CO 80309;
  julie.comerford@colorado.edu}

\altaffiltext{4}{Astronmisches Institut, Ruhr-Universitat Bochum,
  Universitatsstr. 150, 44801, Bochum, Germany;
  middelberg@astro.rub.de}

\begin{abstract} 
  COSMOS\,J100043.15+020637.2 is a merger remnant at $z = 0.36$ with
  two optical nuclei, NW and SE, offset by 500 mas (2.5 kpc).  Prior
  studies suggest two competing scenarios for these nuclei: (1) SE is
  an active galactic nucleus (AGN) lost from NW due to a
  gravitational-wave recoil.  (2) NW and SE each contain an AGN,
  signaling a gravitational-slingshot recoil or inspiralling AGNs.  We
  present new images from the Very Large Array (VLA) at a frequency
  $\nu$ = 9.0 GHz and a FWHM resolution $\theta$ = 320 mas (1.6 kpc),
  and the Very Long Baseline Array (VLBA) at $\nu$ = 1.52 GHz and
  $\theta$ = 15 mas (75 pc).  The VLA imaging is sensitive to emission
  driven by AGNs and/or star formation, while the VLBA imaging is
  sensitive only to AGN-driven emission.  No radio emission is
  detected at these frequencies.  Folding in prior results, we find:
  (a) The properties of SE and its adjacent X-ray feature resemble
  those of the unobscured AGN in NGC\,4151, albeit with a much higher
  narrow emission-line luminosity.  (b) The properties of NW are
  consistent with it hosting a Compton-thick AGN that warms ambient
  dust, photoionizes narrow emission-line gas and is free-free
  absorbed by that gas.  Finding (a) is consistent with scenarios (1)
  and (2).  Finding (b) weakens the case for scenario (1) and
  strengthens the case for scenario (2).  Follow-up observations are
  suggested.
\end{abstract}


\section{Motivation}

Simulations suggest that galaxy mergers can produce remnants with two
or more massive black holes \citep{hof07,ama10,kul12}.  When these
black holes accrete, they can appear as two or more active galactic
nuclei (AGNs) on kiloparsec scales \citep{van12,ble13b}.  Systematic
surveys for such multiple AGNs could provide observational constraints
on AGN activation and tidally enhanced star formation
\citep[e.g.,][]{liu12,kos12}, and on the black hole merger rate, with
consequences for the signals expected for pulsar timing arrays and
gravity-wave detectors \citep{hob10,dot12}.  Guided by early
serendipitous discoveries of dual AGNs like 3C\,75 \citep{owe85} and
NGC\,6240 \citep{bes01,kom03,gal04}, recent systematic surveys are now
yielding spectroscopic samples of dual AGN candidates
\citep[e.g.,][]{com09a,wan09,liu10,smi10,kos12,bar13,com13}.

Candidate dual AGNs can be difficult to confirm due to obscuration and
resolution issues \citep{com11,fu11a,she11,fu12,com12}.  Moreover,
some candidates may turn out to be imposters because they harbor a
recoiling AGN, a situation that both complicates and enriches matters
\citep[e.g.,][]{ble11,gue11,era12}.  This paper focuses on one such
case, COSMOS\,J100043.15+020637.2 (J1000+0206 hereafter).  Early
reports on J1000+0206 \citep{smo08,elv09,com09b,civ10} were prompted
by its unusual optical nature, involving two apparent nuclei in a
merger remnant with a tidal tail \citep{sco07}.

The integrated optical emission-line redshift of J1000+0206 is
$z=0.36$ \citep{lil07}.  Adopting the nomenclature and
cosmology\footnote{$H_0$ = 70 km s$^{-1}$ Mpc$^{-1}$, $\Omega_M$ =
  0.3, $\Omega_\Lambda$ = 0.7} of \citet{civ12}, the two nuclei, SE
and NW, have a projected separation of 500 mas (2.5 kpc).  Two
scenarios have emerged for these optical nuclei: either NW and SE each
contain an AGN with its own narrow emission-line region, or SE is an
AGN recoiling from NW due to the asymmetric emission of gravitational
waves during black-hole coalescence.  Hydrodynamic simulations, with
radiative transfer calculations, of a gas-rich major merger show that
each scenario is consistent with available data \citep{ble13a}.

\citet{smo08} assumed that the emission from J1000+0206 at a frequency
$\nu$ = 1.4 GHz \citep{sch07} is driven by its
spectroscopically-identified AGN.  But this is far from certain.
Figure 1 shows that the radio emission is localized to the inner
portions of the merger remnant \citep{sch07,sch10}, where simulations
indicate that star formation can occur \citep{ble13a}.  Also,
\citet{civ12} recently discovered an extended X-ray feature offset by
about 500 mas (2.5 kpc) to the southwest (SW) of the SE optical
nucleus, further complicating the observational picture.

If the radio emission from J1000+0206 arises from the AGNs and star
formation, Figure 1 makes it clear that disentangling these
contributions requires imaging at subarcsecond resolution.
\S~\ref{imaging} presents new images from the Karl G. Jansky Very
Large Array \citep[VLA;][]{per11} at $\nu$ = 9.0 GHz and a FWHM
resolution $\theta$ = 320 mas (1.6 kpc), and from the Very Long
Baseline Array \citep[VLBA;][]{nap94} at $\nu$ = 1.5 GHz and $\theta$
= 15 mas (75 pc).  \S~\ref{implications} explores the implications of
the new imaging, while a summary and conclusions appear in
\S~\ref{summary}.

\begin{figure}
\plotone{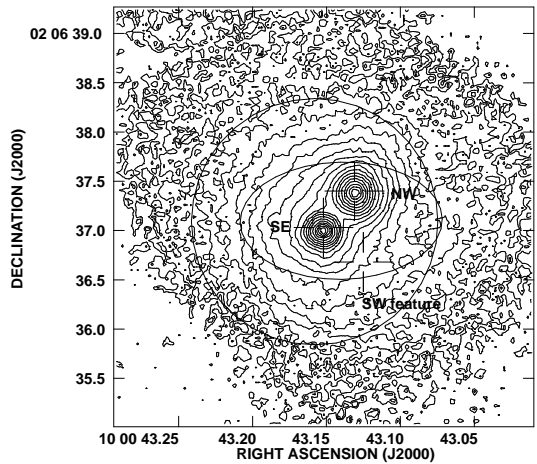}
\caption{HST/ACS image of the F814W emission from J1000+0206 spanning
  4.2\arcsec.  The plus signs mark the positions and 3 $\sigma$
  astrometric errors of the SE and NW optical nuclei (Scoville et
  al.\ 2007), and the approximate position of the X-ray feature offset
  by about 500 mas to the SW of SE (Civano et al.\ 2012).  The circle
  and ellipse are derived from VLA images at $\nu$ = 1.4 GHz with
  $\theta$ = 2.5\arcsec\, and 1.5\arcsec, respectively (Schinnerer et
  al.\ 2010).  A point-like source was detected in the former image,
  and the circle shows its localization as $\theta$ = 2.5\arcsec.  A
  resolved source was detected in the latter image, and the ellipse
  shows its deconvolved Gaussian extent at FWHM of 2.05\arcsec\,
  $\times$ 1.19\arcsec\, with an elongation position angle PA =
  91\arcdeg.  Scale is 1\arcsec\, = 5.03 kpc.}\label{fig1}
\end{figure}

\section{Imaging}\label{imaging}

\subsection{VLA}

J1000+0206 was observed with the VLA in its A configuration on 2012
October 12 UT under proposal code 12B-072.  A coordinate equinox of
2000 was used.  J1024-0052, with a one-dimensional position error of
10 mas (1 $\sigma$), was employed as a phase calibrator.  The
switching time between it and J1000+0206 was 480~s, and involved a
switching angle of 6.6\arcdeg.  The {\em a priori\/} pointing position
for J1000+0206 was shifted 0.6\arcsec\, south of the SDSS position
\citep{com09b} to avoid the risk of phase-center artifacts.  Every 1 s
the correlator produced 1024 contiguous 2-MHz channels, yielding a
total bandwidth of 2.048~GHz per circular polarization centered at
$\nu$ = 9.0~GHz.  Observations of 3C\,138 were used to set the
amplitude scale to an accuracy of about 1\% \citep{per13}.  The net
exposure time on J1000+0206 was about 2150~s.

Release 4.1.0 of the Common Astronomy Software Applications (CASA)
package \citep{mcm07} was used to calibrate and edit the data in an
automated fashion
\footnote{science.nrao.edu/facilities/vla/data-processing/pipeline}.
After further minor edits, the CASA task {\tt imagr} was used to form
and deconvolve a naturally-weighted image of the Stokes $I\/$ emission
from J1000+0206.  This imaged spans 2561 $\times$ 60 mas in each
coordinate, has an rms noise level $\sigma = 0.006$~mJy~beam$^{-1}$
and a geometric resolution at FWHM $\theta$ = 320 mas.  Its inner
portions are shown in Figure 2.  With adequate signal-to-noise,
structures as large as about 10 $\theta$ $\sim$ 3.2\arcsec\, could be
represented in Figure 2.  No emission was detected above a threshold
of $0.018$~mJy~beam$^{-1}$ (3 $\sigma$).

\begin{figure}
\plotone{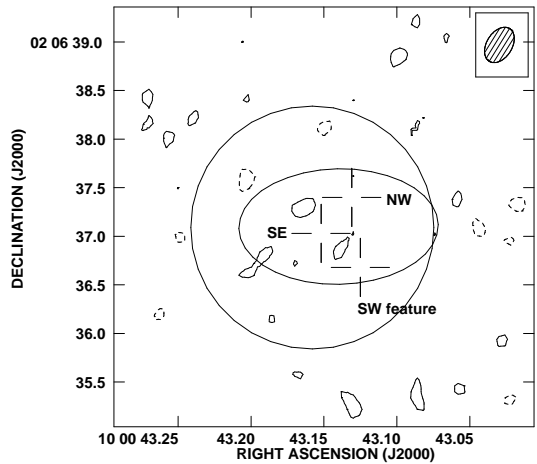}

\caption{VLA image of Stokes $I\/$ emission from J1000+0206 at $\nu$ =
  9.0 GHz and spanning 4.2\arcsec.  The rms noise is
  0.006~mJy~beam$^{-1}$ (1 $\sigma$) and the beam dimensions at FWHM
  are 400 mas $\times$ 260 mas with an elongation PA = -32\arcdeg\,
  (boxed hatched ellipse).  Contours are at -6, -4, -2, 2, 4, and 6
  times 1 $\sigma$.  Negative contours are dashed and positive ones
  are solid.  No emission was detected above 3 $\sigma$.  The circle,
  ellipse, plus signs and scale are the same as for
  Fig.~1.}\label{fig2}
\end{figure}

\subsection{VLBA}

J1000+0206 was observed with the VLBA during four 6-hour sessions
spanning 2012 June 24 to 2013 January 11 under proposal code BM360.  A
coordinate equinox of 2000 was used.  J1011+0106, with a
one-dimensional position error of 2 mas (1 $\sigma$), was used as a
phase calibrator.  The switching time between it and J1000+0206 was
360~s, and involved a switching angle of 2.7\arcdeg.  Every 4.1 s the
correlator \citep{del11} produced 128 contiguous 2-MHz channels,
yielding a total bandwidth of 0.256~GHz per circular polarization
centered at $\nu$ = 1.52~GHz.  VLBA system temperatures and gains were
used to set the amplitude scale to an accuracy of about 5\%.  A total
of about 450 baseline hours were accrued on J1000+0206.

For each 6-hour session, NRAO's Astronomical Image Processing System
\citep[AIPS;][]{gre03} was used to calibrate and edit the data
following the approach described by \citet{mid13}.  The AIPS task {\tt
  imagr} was used to form and deconvolve a naturally-weighted image of
the Stokes $I\/$ emission from J1000+0206.  This image spans 2048
$\times$ 1 mas in each coordinate, has a geometric resolution at FWHM
$\theta$ = 15 mas, and was corrected for primary beam attenuation.
Combining the u-v data from the four sessions resulted in an rms noise
level $\sigma = 0.013$~mJy~beam$^{-1}$.  With adequate
signal-to-noise, structures as large as about 10 $\theta$ $\sim$ 150
mas could be imaged.  No emission was detected above a threshold of
$0.078$~mJy~beam$^{-1}$ (6 $\sigma$).

\section{Implications}\label{implications}

\subsection{Overview}\label{overview}

The new VLBA image at $\nu$ = 1.52 GHz and $\theta$ = 15 mas (75 pc)
filters for emission with a rest-frame brightness temperature $T_b >
3.5 \times 10^5$~K, not achieved by even the most compact starbursts
\citep{con92}.  Moreover, the VLBA image is too shallow to detect even
the most luminous radio supernova beyond $z \sim 0.1$
\citep[e.g.,][]{gar05,mid13}.  Thus the VLBA image is insensitive to
emission driven by star formation.  In contrast, the new VLA image
(Figure 2) at $\nu$ = 9.0 GHz and $\theta$ = 320 mas (1.6 kpc) is
sensitive to emission driven by star formation, and to AGN-driven
emission in the SE or NW optical nuclei or associated with the X-ray
feature offset to the SW of the SE optical nucleus
\citep{com09b,civ10,civ12}.

Applying a K-correction for a spectral index $\alpha = -0.7$ ($S
\propto \nu^\alpha$), the VLA nondetections in Figure 2 imply that on
scales of 1~$\theta$ = 0.32\arcsec\, (1.6 kpc) to 10~$\theta$ =
3.2\arcsec\, (16 kpc) the merger remnant has a spectral luminosity
$L_{\rm 9.0~GHz} < 7.3 \times 10^{21}$ W Hz$^{-1}$.  Given these
sub-galactic scales, we adopt local indicators of star formation rates
(SFRs) \citep[e.g.,][]{cal12}.  We apply equation (15) of
\citet{mur11} at a rest-frame frequency $\nu = 9.0(1+z)$~GHz =
12.2~GHz, include a thermal contribution for an electron temperature
$T_e = 10^4$~K and a nonthermal contribution with a spectral index
$\alpha = -0.8$, and find a SFR $<$ 17 M$_\odot$ yr$^{-1}$ for a
Kroupa initial mass function (IMF).  This is an interesting regime to
be exploring, given that the host galaxies of optically-selected AGNs
at $z < 0.3$ could have SFR $\sim$ 1-20 M$_\odot$ yr$^{-1}$, spanning
values for normal spiral galaxies to those for moderately luminous
starbursts \citep{kim06,con13}.

Prior VLA imaging at $\nu$ = 1.4 GHz \citep{sch10} localized the
emission to the circular and elliptical regions, outlined in Figures 1
and 2, that encompass the inner merger remnant, the optical nuclei SE
and NW, and the X-ray feature SW of the SE nucleus.  The circular
region marks a point-like source with an integrated flux density
$S_{\rm 1.4~GHz} = 0.113 \pm 0.010$ mJy and a K-corrected spectral
luminosity $L_{\rm 1.4~GHz} \sim 4.6 \times 10^{22}$ W Hz$^{-1}$.  If
driven by star formation only, the corresponding SFR = 34 M$_\odot$
yr$^{-1}$ \citep[][eqn.\ 15]{mur11}.  This is at least a factor of two
higher than the new VLA limit derived above, meaning that up to half
of the integrated flux density could arise from star formation.
Conversely, any AGN-driven emission could have an integrated $S_{\rm
  1.4~GHz} \sim 0.056-0.113$ mJy and a K-corrected luminosity $\nu
L_{\nu}(1.4~\rm{GHz}) = (3-6) \times 10^{38}$ erg s$^{-1}$, the regime
of low-luminosity AGN (LLAGN) at $z=0$ \citep{ho08}.

Analyses of the {\it HST} F814W ACS image and the Keck/DEIMOS spectrum
of the galaxy were reported in \citet{com09b}, and can be summarized
as follows.  Using Source EXtractor on the {\it HST} image, the SE and
NW optical nuclei were found to be aligned along a PA = 139.6\arcdeg,
with a projected separation of $500 \pm 9$ mas ($2.50 \pm 0.05$ kpc).
The DEIMOS spectrum was taken with a slit PA = 143.3\arcdeg\, and
spanning 4730--9840 \AA \, in wavelength.  This spectrum revealed
double narrow emission-line components, with line flux ratios of both
components falling within the AGN region of the
Baldwin-Phillips-Terlevich diagram, within the 1 $\sigma$ errors on
the flux measurements.

Projected spatial separations between the double emission-line
components of $2.1 \pm 0.7$ kpc, $1.4 \pm 0.6$ kpc, and $2.3 \pm 0.6$
kpc were measured for \mbox{[\ion{O}{3}] $\lambda$5007},
\mbox{H$\alpha$}, and \mbox{[\ion{N}{2}] $\lambda$6584}, respectively.
The mean of the three spatial separations measurements, weighted by
their inverse variances, is $1.9 \pm 0.4$ kpc.  However, the
\mbox{H$\alpha$} spatial separation measurement is not consistent with
the separations measured in \mbox{[\ion{O}{3}] $\lambda$5007} and
\mbox{[\ion{N}{2}] $\lambda$6584} and is skewed low because of partial
obscuration by an imperfectly subtracted night-sky line.  Using only
the spatial separations measured in \mbox{[\ion{O}{3}] $\lambda$5007}
and \mbox{[\ion{N}{2}] $\lambda$6584} (which are consistent with one
another, to within the 1 $\sigma$ errors), the
inverse-variance-weighted mean spatial separation is $2.2 \pm 0.5$
kpc.

In projection, the spatial separation between the double emission-line
components ($2.2 \pm 0.5$ kpc) is consistent with the spatial
separation between the double nuclei in the {\it HST} image ($2.50 \pm
0.05$ kpc), to within their errors.  There is a 3.7\arcdeg\,
difference between the PA of the DEIMOS slit and the PA of the two
nuclei in the {\it HST} image, but this results in a negligible
(0.2\%) difference between the spectrum-based and image-based
measurements of spatial separations.  These similar spatial
separations, combined with the excitation state of the narrow
emission-line components, are taken as strong indicators for the
presence of two AGNs, likely associated with SE and NW.

The exposure time of the DEIMOS spectrum was too short to allow us to
localize, with confidence, each narrow emission-line component with
each optical nucleus.  For now we make the plausible assumption that
\mbox{[\ion{O}{3}] $\lambda$5007} luminosities $L({\rm [OIII]}) =
4.8-9.9 \times 10^{42}$ erg s$^{-1}$ and excitation ratios
[OIII]$\lambda$5007/H$\beta$ $= 3.4-4.4$ for the two AGNs
\citep{com09b}.  In the near future we will obtain Keck/OSIRIS data to
test this important assumption (F.\, Muller-Sanchez, priv.\, comm.).

\subsection{The SE Optical Nucleus and its SW X-ray Feature}\label{se}

\citet{civ10,civ12} provide strong evidence that the SE optical
nucleus is an unobscured AGN of Type 1, thus featuring a broad
emission-line region.  SE has an X-ray counterpart with a luminosity
$L(2-10~\rm{keV}) = 1.14 \times 10^{43}$ erg s$^{-1}$
\citep{civ10,civ12}.  Its Eddington ratio is 0.04 \citep{civ10,tru11},
typical for a Type 1 AGN in COSMOS \citep{tru09}.  The new VLA and
VLBA nondetections of SE imply that any AGN-driven emission has
K-corrected luminosities $\nu L_{\nu}(9.0~\rm{GHz}) < 6.6 \times
10^{38}$ erg s$^{-1}$ (3 $\sigma$) on scales below 1.6 kpc and $\nu
L_{\nu}(1.52~\rm{GHz}) < 4.7 \times 10^{38}$ ergs s$^{-1}$ (6
$\sigma$) on scales below 75 pc.  From \S~\ref{overview}, the
luminosity of any AGN-driven emission from SE cannot exceed $\nu
L_{\nu}(1.4~\rm{GHz}) = 6 \times 10^{38}$ erg s$^{-1}$.

Following \citet{ter03}, the new VLA photometry for SE yields log $\nu
L_{\nu}(9.0~\rm{GHz}) / L(2-10~\rm{keV}) < -4.2$.  This value is
consistent with SE's modest Eddington ratio \citep{ho08}.  From
\S~\ref{overview}, the \mbox{[\ion{O}{3}] $\lambda$5007} luminosity of
SE is $L({\rm [OIII]}) = 4.8-9.9 \times 10^{42}$ erg s$^{-1}$
\citep{com09b}.  The VLBA luminosity, free from contamination from
star formation, thus implies that SE is radio quiet as defined by
\citet{zak04} or \citet{tru11,tru13}, a trait that is also consistent
with its modest Eddington ratio.

For context, several key properties of SE resemble or are consistent
with those of the prototypical Type 1 AGN NGC\,4151.  Adopting a
distance of 13.3 Mpc for NGC\,4151, it has (a) $L(2-10~\rm{keV}) = 5.0
\times 10^{42}$ erg s$^{-1}$ \citep{ho09}; (b) when integrated within
a radius of 2\arcsec\, (130 pc), $\nu L_{\nu}(1.425~\rm{GHz}) = 1.0
\times 10^{38}$ erg s$^{-1}$, $\nu L_{\nu}(8.465~\rm{GHz}) = 1.4
\times 10^{38}$ erg s$^{-1}$ and $\alpha = -0.75$ \citep{kuk95,ho01};
and (c) log $\nu L_{\nu}(8.465~\rm{GHz}) / L(2-10~\rm{keV}) = -4.6$.
\citet{mun03} and \citet{wan11a} report evidence that the radio source
dynamically interacts with the interstellar medium (ISM) at these
radii in NGC\,4151.

\citet{civ12} noted that the X-ray feature offset by about 500 mas
(2.5 kpc) to the SW of the SE optical nucleus could be driven by star
formation.  Indeed, \citet{del08} suggest that the impulse from the
passage of a recoiling black hole could could trigger star formation
in its wake.  In the scenario where SE has recoiled from NW, the SW
X-ray feature might be its star formation wake.  That feature is about
18\% as luminous as SE in the passband of the {\em Chandra\/} High
Resolution Camera (HRC), indicating that it has $L(2-10~\rm{keV}) =
2.1 \times 10^{42}$ erg s$^{-1}$ \citep{civ10,civ12}.

From \S~\ref{overview}, the new VLA upper limit to the SFR at the
location of the SW X-ray feature is SFR $<$ 17 M$_\odot$ yr$^{-1}$ for
a Kroupa IMF.  This corresponds to $L(2-10~\rm{keV}) < 8.5 \times
10^{40}$ erg s$^{-1}$ according to \citet{ran03}, who use a Salpeter
IMF.  Even with this caveat of different IMFs, the X-ray emission from
the SW feature is much too luminous to be driven by star formation.
This points to the SW X-ray feature being AGN-driven, perhaps, as
\citet{civ12} suggest, because SE photoionizes the ambient ISM on
kiloparsec scales and produces optical and X-ray emission lines
analogous to the situation at radii beyond 2\arcsec\, (130 pc) in
NGC\,4151 \citep{wan11b}.  Notably, SE's [OIII]$\lambda$5007
luminosity is about 37-76 times that of NGC\,4151 on kpc scales
\citep{ho01}, perhaps because of SE's location in a merger remnant
rather than in a typical LLAGN host.

\subsection{The NW Optical Nucleus}\label{nw}

The new VLA and VLBA nondetections of NW imply that any AGN-driven
emission has K-corrected luminosities $\nu L_{\nu}(9.0~\rm{GHz}) < 6.6
\times 10^{38}$ erg s$^{-1}$ (3 $\sigma$) on scales below 1.6 kpc and
$\nu L_{\nu}(1.52~\rm{GHz}) < 4.7 \times 10^{38}$ ergs s$^{-1}$ (6
$\sigma$) on scales below 75 pc.  From \S~\ref{overview}, the
luminosity of any AGN-driven emission from NW cannot exceed $\nu
L_{\nu}(1.4~\rm{GHz}) = 6 \times 10^{38}$ erg s$^{-1}$.  From
\S~\ref{overview}, the \mbox{[\ion{O}{3}] $\lambda$5007} luminosity of
NW is $L({\rm [OIII]}) = 4.8-9.9 \times 10^{42}$ erg s$^{-1}$
\citep{com09b}.  The VLBA luminosity, free from contamination from
star formation, thus implies that NW is radio quiet following the
\citet{zak04} definition.  The X-ray emission from the NW optical
nucleus is less than 4.5\% as luminous as SE in the HRC passband,
suggesting that NW has $L(2-10~\rm{keV}) < 5.1 \times 10^{41}$ erg
s$^{-1}$ \citep{civ10,civ12}.

The NW optical nucleus could harbor an intrinsically faint AGN, with a
narrow emission-line region only.  However, evidence for a very
obscured Type 2 AGN in NW is beginning to emerge.  Recent
prescriptions for identifying obscured AGNs \citep{jun11} can be
applied.  From \S~\ref{overview}, the excitation ratio of NW is
[OIII]$\lambda$5007/H$\beta$ $= 3.4-4.4$ \citep{com09b}.  Such a
ratio, when combined with estimates for the host galaxy's stellar mass
$M_\star \sim 10^{10-11} M_\odot$ \citep{civ10,kar10}, occupy the AGN
locus of the mass-excitation diagram \citep[Fig.~4;][]{jun11}.
Moreover, as \citet{civ12} suggest, NW could lack an X-ray counterpart
due to atomic absorption.  Comparing NW's Compton-thickness parameter
$T = L(2-10~\rm{keV}) / L({\rm [OIII]})$ to its \mbox{[\ion{O}{3}]
  $\lambda$5007} luminosity \citep{com09b,civ12} identifies it as a
Compton-thick candidate \citep[Fig.~10;][]{jun11}.

NW could also lack a counterpart at $\nu$ = 9.0 GHz due to free-free
absorption by its narrow emission-line gas, a concept applied to
account for the flat radio spectra of some Type 2 AGN with $L({\rm
  [OIII]}) = 10^{42}$ erg s$^{-1}$ \citep{lal10}.  For a
characteristic electron temperature $T_e = 10^4$~K and density $n_e =
10^3$~cm$^{-3}$, a path length of about 380 pc gives a free-free
optical depth of unity at $\nu$ = 10~GHz \citep{ulv01}.  Intriguingly,
this free-free path length resembles the scale length of about 100 mas
(500 pc) reported for NW's starlight \citep{civ10}.  This path length
can be cross-checked in several ways.  First, the H$\beta$ luminosity
for either the blueshifted or redshifted narrow emission-line gas
\citep{com09b} shouldn't be exceeded; this can be achieved by invoking
a plausible volume-filling factor $f \sim 1-3 \times 10^{-3}$.
Second, if the narrow emission-line gas has a spherical geometry with
diameter 150 mas (2 $\times$ 380 pc), the source at $\nu$ = 9.0 GHz
should be much smaller; this can be checked with sensitive VLBA
imaging.  Third, at VLBA resolutions the spectrum of the source at
frequencies $\nu <$ 9.0 GHz should be exponentially suppressed; this
is also testable with sensitive VLBA imaging.

\citet{civ12} assembled the spectral energy distribution at
far-infrared and shorter wavelengths, corrected it for the
contributions of the Type 1 AGN in SE, and inferred an infrared (IR)
luminosity $L({\rm IR}) \sim 6 \times 10^{44}$ erg s$^{-1}$.  That
luminosity corresponds to a Kroupa-based SFR $\sim$ 23 M$_\odot$
yr$^{-1}$ \citep[][eqn.\ 4]{mur11}, although this value would be
overestimated if dust warmed by an obscured Type 2 AGN in NW
contributed to the IR luminosity \citep[e.g.,][]{deg92,jun13}.  From
\S~\ref{overview}, the new VLA upper limit to the SFR at the location
of the NW optical nucleus is SFR $<$ 17 M$_\odot$ yr$^{-1}$.  This
lower radio-based SFR suggests that the IR photometry could suffer
some contamination from dust warmed by an obscured AGN.

\section{Summary and Conclusions}\label{summary}

We presented new radio imaging of J1000+0206, a merger remnant at $z =
0.36$ with two optical nuclei, NW and SE, offset by 500 mas (2.5 kpc).
The new VLA imaging at $\nu$ = 9.0 GHz is sensitive to emission driven
by AGNs and/or star formation, and the new VLBA imaging at $\nu$ =
1.52 GHz is sensitive only to emission driven by AGNs.  No radio
emission was detected at these frequencies.  Combining new and prior
results, the following self-consistent picture emerged:

\begin{itemize}

\item The new VLA photometry at $\nu = 9.0$~GHz implies a SFR $<$ 17
  M$_\odot$ yr$^{-1}$ for the inner merger remnant, the optical nuclei
  SE and NW, and the X-ray feature adjacent to the SE nucleus.  After
  correcting the prior VLA photometry at $\nu = 1.4$~GHz for this SFR,
  the AGN-driven emission has an integrated luminosity $\nu
  L_{\nu}(1.4~\rm{GHz}) = (3-6) \times 10^{38}$ erg s$^{-1}$, in the
  realm of LLAGN in the local Universe.

\item The properties of SE and its adjacent X-ray feature match those
  of the unobscured prototype AGN in NGC\,4151.  However, the
  luminosity of the narrow emission-line gas associated with SE is
  about forty to eighty times higher than that associated with the AGN
  in NGC\,4151.  Deeper radio, optical and X-ray studies of SE will
  test its resemblance to, and difference from, the Type 1 AGN
  prototype in NGC\,4151.  For example, the difference in narrow
  emission-line luminosities could reflect atypical, merger-induced
  conditions near SE.

\item The properties of NW are consistent with it hosting a
  Compton-thick AGN that warms the ambient dust and photoionizes the
  narrow emission-line gas, and is free-free absorbed by that gas at
  $\nu$ = 9.0 GHz.  These findings enhance the prospects that
  AGN-driven emission could emerge from radio and X-ray studies of NW
  at deeper levels and X-ray studies of NW at harder wavelengths.  For
  example, deeper VLBA imaging could reveal a radio spectrum that is
  exponentially suppressed by free-free absorption. These findings
  also underscore the urgency of localizing narrow emission-line gas,
  whether blueshifted or redshifted, to NW.

\end{itemize}

In the scenario where SE is an AGN that has recoiled from NW due to
the asymmetric emission of gravitational waves during black-hole
coalescence, NW cannot host an obscured AGN.  The emerging evidence
for a very obscured Type 2 AGN in NW thus both weakens the case for SE
being a gravitational-wave recoil and strengthens the case that each
optical nucleus contains it own AGN.  Two interpretations are
consistent with the latter scenario: (i) SE and NW mark inspiralling
dual AGNs like those sought in systematic spectroscopic surveys for
dual AGN candidates
\citep{com09a,wan09,liu10,smi10,kos12,bar13,com13}, or (ii) SE is a
gravitational-slingshot recoil moving away from NW and the Type 2 AGN
that it harbors \citep{civ10,civ12,ble13a}.

It must be noted that SE's broad H$\beta$ emission line is redshifted
by more than 1000 km~s$^{-1}$ relative to the narrow H$\beta$ emission
line \citep{civ10}.  Such large velocity offsets occur in less than 1
\% of normal Type 1 AGN \citep{bon07,era12}, making it very unlikely
that they will occur by chance in an inspiralling Type 1 AGN.  In
contrast, such large velocity offsets are predicted for a
gravitational-slingshot recoil \citep{hof07}, a point in favor of that
interpretation for J1000+0206.

\acknowledgements We thank the referee for a timely and helpful
report.  We also acknowledge using Ned Wright's Cosmology Calculator
\citep{wri06} and benefiting from helpful discussions with Michael
Cooper, Francisco Muller-Sanchez, Jonathan Trump and Craig Walker.  EM
acknowledges financial support from the Deutsche
Forschungsgemeinschaft through project FOR1254.

{\it Facilities:} \facility{VLA}, \facility{VLBA}.

\end{document}